# Opportunities and Challenges of Applying Large Language Models in Building Energy Efficiency and Decarbonization Studies: An Exploratory Overview

Liang Zhang[1,2,*], Zhelun Chen[3]
1. University of Arizona, 3740 E 34th St, Tucson, AZ 85721, US
2. National Renewable Energy Laboratory, 15013 Denver West Parkway, Golden, CO 80401, US
3. Drexel University, Philadelphia, 3141 Chestnut St, Philadelphia, PA 19104, US

## Abstract

In recent years, the rapid advancement and impressive capabilities of Large Language Models (LLMs) have been evident across various domains. This paper explores the application, implications, and potential of LLMs in building energy efficiency and decarbonization studies. The wide-ranging capabilities of LLMs are examined in the context of the building energy field, including intelligent control systems, code generation, data infrastructure, knowledge extraction, and education. Despite the promising potential of LLMs, challenges including complex and expensive computation, data privacy, security and copyright, complexity in fine-tuned LLMs, and self-consistency are discussed. The paper concludes with a call for future research focused on the enhancement of LLMs for domain-specific tasks, multi-modal LLMs, and collaborative research between AI and energy experts.

## Keywords

## List of Abbreviations

ACC: automated compliance checking

AI: artificial intelligence

IML: interpretable machine learning

LLM: large language model

MLC: machine learning control

NLP: natural language processing

RT-2: Robotic Transformer 2

VLA: vision-language-action

## 1 Introduction

Buildings consume a significant portion of global energy production and are a major contributor to greenhouse gas emissions. According to the U.S. Energy Information Administration, in 2022, the building sector was responsible for 29% of the total end-use energy utilization in the U.S.; factoring in the energy losses from the electrical systems within this sector, it represented 40% of the entire U.S. energy usage [1]. Enhancing energy efficiency in buildings can notably decrease energy use and related emissions, making it a key strategy in combating climate change. Decarbonization, which involves reducing carbon dioxide and other greenhouse gas emissions, is another critical component of sustainability efforts in the building sector. Strategies for decarbonization include shifting to renewable energy solutions, boosting energy-saving practices, and utilizing carbon capture and storage technologies. Improving building energy efficiency and decarbonizing buildings are inextricably linked, each embedding element of the other.

---

[*] Corresponding Author: liangzhang1@arizona.edu

Building energy efficiency and decarbonization studies are multidisciplinary fields that combine aspects of architectural engineering, mechanical systems design and control, environmental science, urban science, and policy development to develop and implement solutions that can reduce the environmental impact of buildings. Current building energy efficiency and decarbonization studies involve but are not limited to energy auditing, simulation and modeling, energy-saving technologies such as insulation, energy-efficient appliances, energy management systems, use of renewable energy sources, building electrification, carbon capture, and storage technologies, and the development of zero-energy buildings.

Nevertheless, as building energy studies grow in complexity, we identify pivotal challenges in their development, particularly pertaining to the escalating need for automation in managing the exponential surge in text and data associated with decarbonization research. The challenges include, but are not limited to, the following. **Imperative for Robust Automation**. In the context of building energy studies, handling vast and multifaceted datasets alongside modeling intricate energy systems is a common requirement. Automation emerges as a critical factor, streamlining analyses and ensuring results are obtained more swiftly, consistently, and accurately. This enhancement in efficiency and reliability is indispensable for advancing research in the field. **Challenges in Knowledge Extraction**. A wealth of textual data, encompassing building energy standards, energy policy documents, research papers, and technical reports holds invaluable insights. However, the task of extracting and structuring this information is intricate and demands substantial resources, posing a significant challenge. **Evolving Educational Needs.** The escalating complexity of building energy studies highlights the pressing need for more comprehensive education and training for building and urban scientists. However, existing training methodologies, predominantly rooted in traditional resources such as textbooks, papers, and reports, may fall short in addressing these evolving needs. This gap underscores the difficulty for learners in acquiring the specialized training they necessitate.

The rapid progression in the field of Artificial Intelligence (AI) has facilitated the emergence of Large Language Models (LLMs) like ChatGPT, offering potential applications extending into various fields including building energy efficiency and decarbonization studies. LLMs can be a good solution to the challenges mentioned earlier. They can process and analyze large volumes of data to provide insights, assist in decision-making, extract valuable information from textual data, and engage the public. They can also be used to develop and improve software and modeling tools, ensure regulatory compliance, and develop educational materials. However, how LLM can practically impact building energy efficiency and decarbonization studies remains unexplored.

The paper opens a discussion on the exploration of the application, implications, and potential of LLMs in building energy efficiency and decarbonization studies. This paper is organized as follows: After the introduction, we present an overview of LLMs, followed by a discussion on the potential applications of

LLMs in building energy efficiency and decarbonization studies. We then discuss the challenges and limitations of using LLMs in these fields and conclude with future research directions.

## 2 Overview of Large Language Models (LLMs)

Large Language Models (LLMs), such as ChatGPT and Llama, are a class of AI models that have shown remarkable capabilities in understanding, generating, and translating human language. These models are trained on great amounts of text data, enabling them to answer questions with human-like text, write essays, summarize documents, and even translate languages. The "large" in denotes both the volume of training data and the quantity of parameters within the models.

The development of LLMs has been a gradual process, starting with smaller models such as Eliza and PARRY in the 1960s and 1970s, to more sophisticated models like ChatGPT and BERT in recent years. The biggest leap in the development of LLMs has been the introduction of Transformer architectures, which use self-attention mechanisms to consider the context of all words in the input when generating an output. The latest versions of these models, like GPT-4, contain hundreds of billions of parameters and are capable of impressive feats of language understanding and generation.

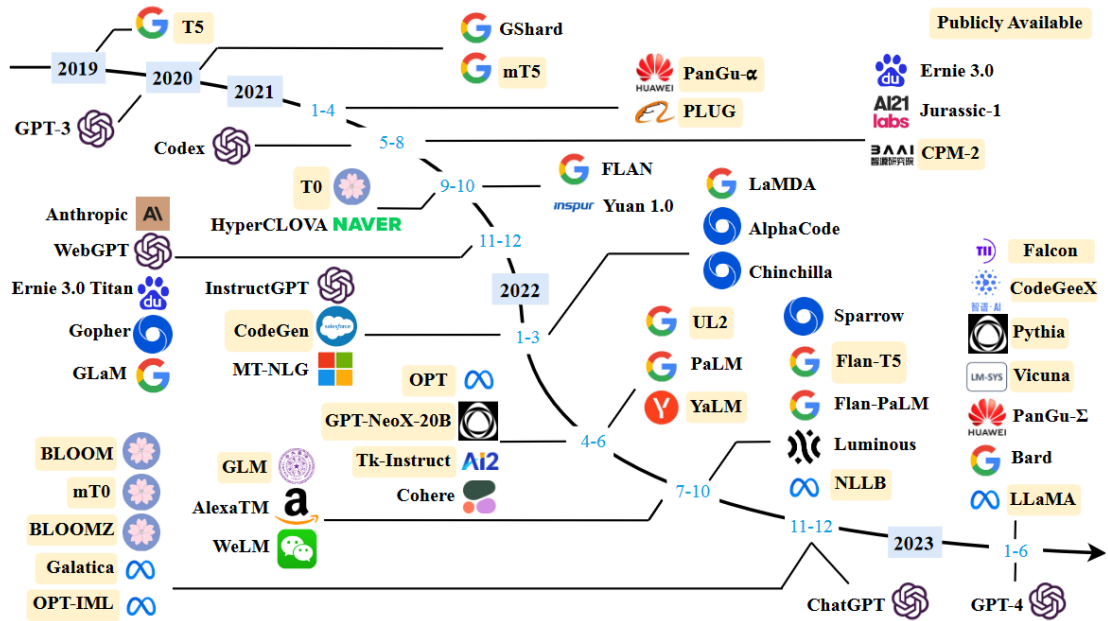

Figure 1. Timeline of existing LLMs (parameter size larger than 10 billion) in recent years [2]

Traditional human-computer interactions have predominantly revolved around humans interfacing with data and information through specific applications (such as software and computer programming). Different types of data, such as structured (like databases or tables) and unstructured (like text, images, videos), require different software. This interaction has historically been complex and diverse. With the advent of LLM, human-computer interactions have fundamentally changed. In the future, instead of interacting with data through specific applications, humans interact directly with an all-in-one interface connecting LLM. Applications have been relegated to the background, and the interaction has become direct and unified. LLM has come to the forefront, with applications hidden in the background. The interaction between humans and data in the era of LLM appears very simple. People can express their requests or commands, and the LLM seamlessly takes over. Beneath this straightforward interface, the LLM is adept at navigating and executing intricate processes, including 1) understanding natural language, 2) planning tasks, and 3) converting to formal language. For complex tasks, it is best to break them down into simpler tasks and solve them one by one. Despite the interaction with humans in natural language, subsequent data processing generally requires formal language, such as programming, APIs, SQL, etc. to be executed in the machines and systems. To summarize, LLM has the great potential to change the way people interact with the world, including improving building energy efficiency and decarbonizing buildings.

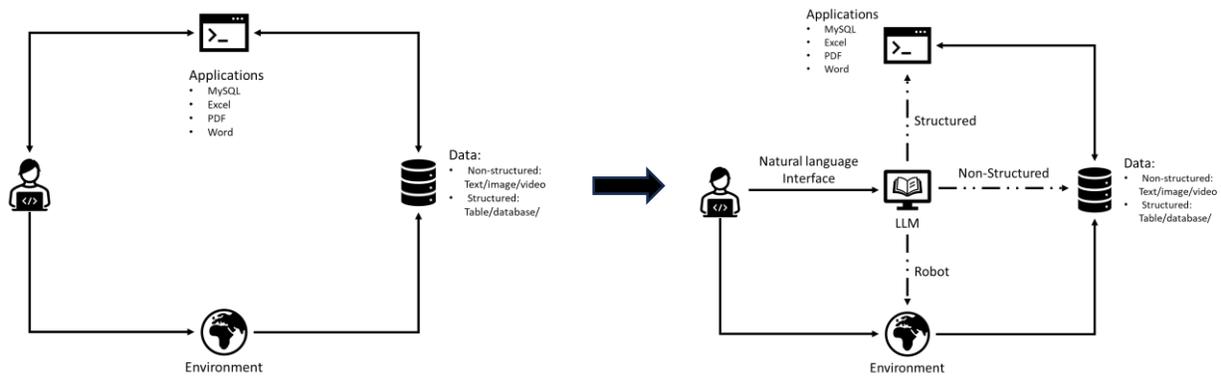

Figure 2. LLM changes the way people interact with data, applications, and the environment.

LLMs already have a wide range of applications across various fields. In healthcare, they can help answer medical queries, assist in patient diagnosis [3], and analyze health record [4]. In education, they can create personalized learning materials, automate grading, and provide tutoring [5]. In business, they can automate customer service (e.g., chatbot [6]), generate marketing content, develop recommender system [7], and provide business intelligence. In the legal field, they can help with contract analysis, legal research, and case prediction [8]. In the technology sector, they can assist in software development, bug fixing, and code generation [9]. In the scientific realm, LLM can store, merge, and rationalize scientific information (e.g., Galactica [10]). In finance, LLMs can analyze market trends, forecast stock movements, automate financial reporting, and provide investment insights. With tools like BloombergGPT [11], they can synthesize vast amounts of financial data into concise reports, offer real-time market updates, and even predict economic shifts based on historical and current data patterns. Despite the diverse applications in various fields, the utilization of LLMs in the niche area of building energy efficiency and decarbonization studies remains largely unexplored. Therefore, Section 3 delves deeper into this specific application.

# 3 Potential Applications of LLMs in Building Energy Efficiency and Decarbonization Studies

The evolution of Natural Language Processing (NLP) has undergone significant transformations over the decades. Initially, NLP systems relied on handcrafted rules and lexicons to parse and understand text. However, the advent of machine learning shifted the paradigm, ushering in the era of pretrained language models that utilized vast amounts of data to understand linguistic patterns. These models, such as Word2Vec [12] and GloVe [13], captured semantic meanings of words in continuous vector spaces. The momentum further accelerated with the introduction of transformer architectures, like BERT and GPT, marking the rise of LLMs. These LLMs, capable of processing vast datasets, have displayed unprecedented proficiency in comprehending and generating human-like text. Their emergence has reshaped the landscape of NLP, offering capabilities far beyond traditional methods and setting new benchmarks across a plethora of language tasks. Figure *3* illustrates the advanced capabilities of LLMs, distinguishing them from conventional pre-trained language models. Features such as few-shot prompting, step-by-step reasoning, in-context learning, and instruction following elevate the potential of LLMs beyond traditional NLP applications.

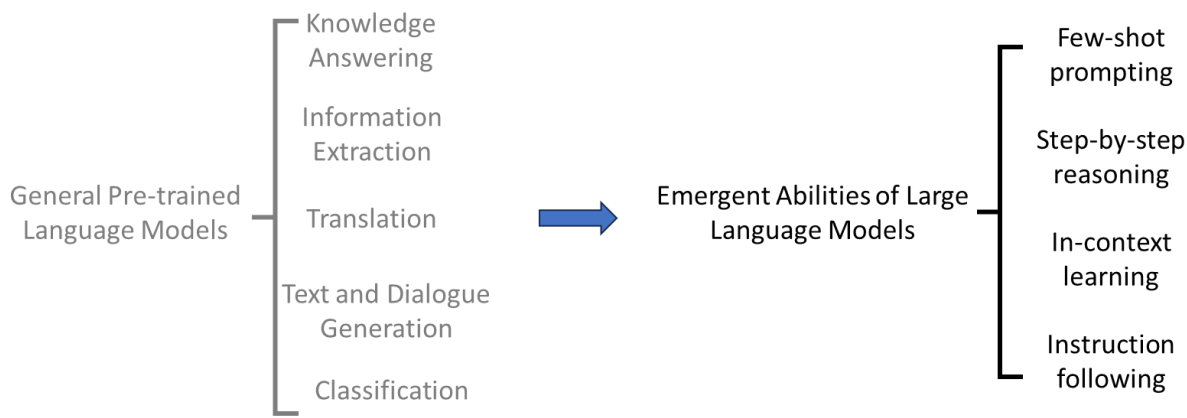

Figure 3. Diagram of emergent abilities of LLMs [2] [14]

Drawing upon the capabilities inherent to NLP and the emergent abilities of LLMs, we have identified and examined seven application areas where the fullest potential of these advanced features can be realized and harnessed to a significant extent.

## 3.1 Intelligent Control Systems

**Control of Building Energy Systems**. LLMs can be an integral part of intelligent control systems in buildings, providing insights that can help optimize building energy use and decarbonization efforts. Envisioning LLMs as agents in the realm of building control offers a transformative approach to intelligent building operations. LLM-as-agent can interpret vast amounts of data from building automation system to access indoor environment, occupant behavior, user inputs and preferences, and external sources like weather data, and subsequently make decisions or provide recommendations in real-time. This agent-like behavior can improve the dynamic adaptability of building systems due to the in-context learning capability of LLMs. This allows the LLM to provide both predictive control, which anticipates future energy needs and adjusts systems in advance, and proactive control, which actively seeks opportunities to optimize energy consumption based on immediate conditions. For instance, in scenarios where there is a sudden change in occupancy or external environmental conditions, LLMs can quickly analyze the situation and adjust control parameters to maintain optimal indoor conditions while ensuring energy efficiency. Song et al. [15] undertook multiple tests to address these inquiries: 1) To what extent is GPT-4 effective in HVAC management? 2) How adept is GPT-4 at adapting to varied HVAC control situations? 3) How do distinct configurations influence the outcome? They found GPT-4 delivers performance on par with reinforcement learning techniques, yet requires fewer samples and has reduced technical complexities, suggesting the feasibility of using LLMs directly for HVAC control assignments. However, this work is limited by using LLM to mimic the pre-collected advanced control strategies, which limits the potential of LLM to outperform existing control algorithms.

**Improved Interaction with Building Energy Management Systems.** Moreover, the conversational ability of LLMs can enable a more intuitive interface for building operators to interact with building energy management system or building automation systems. Instead of navigating through complex and varieties of dashboards and database provided by different vendors, operators can interact with the LLM-agent using natural language, asking questions, querying data, visualization, or issuing commands, making the monitoring and control process more user-friendly. It eliminates the technical obstacles in accessing and operating building, traditionally necessitating specialized training and coding expertise. This also facilitates a two-way communication channel: the LLM can proactively notify operators about potential issues or optimizations and can also receive feedback, refining its operations over time. To illustrate this concept, Figure *4* shows how the user can control building energy systems via LLM-as-agent workflow. The example specifically shows the workflow, from user's request to ensure indoor air quality due to an abnormal occupancy change, to the actual control HVAC systems.

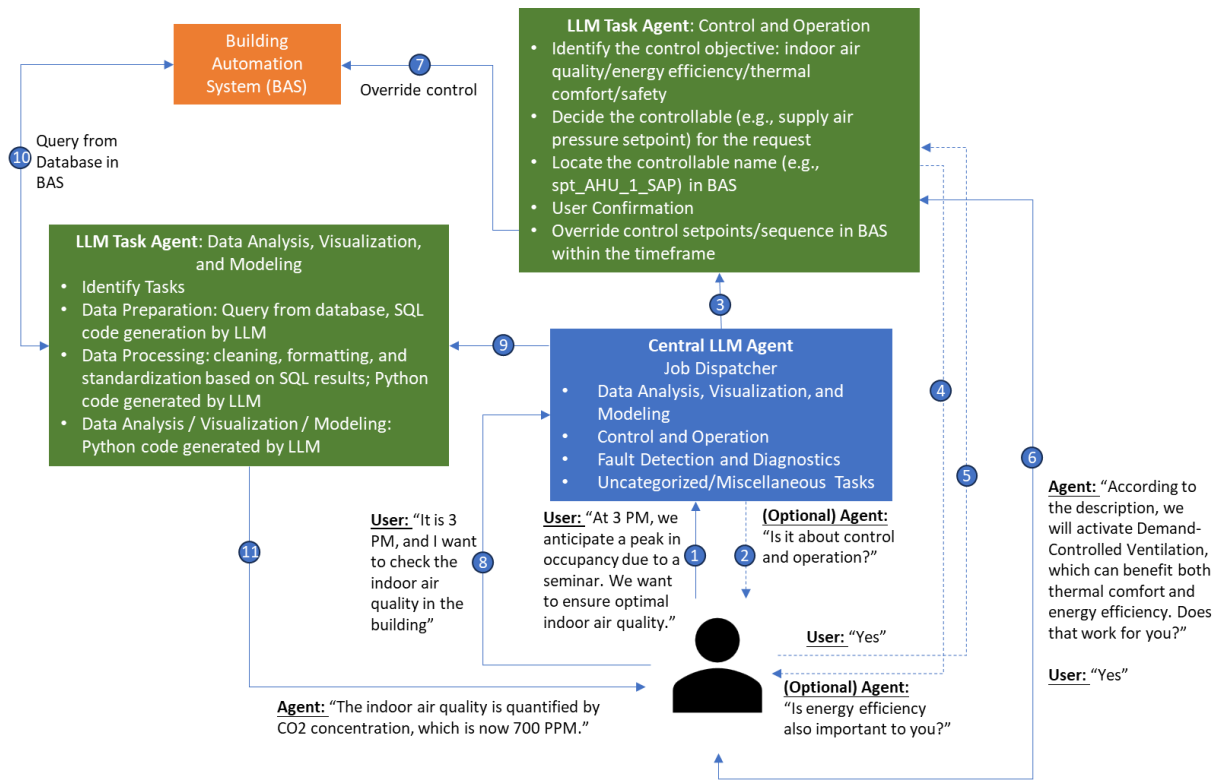

Figure 4. Diagram of LLM-based interaction with building energy management systems with an example

**Built Environment Robots.** LLMs have made significant advancements in robotic control. Liang et al. [16] introduced a concept termed "Code as Policies." This approach revolves around using language models to craft robot-focused programs that embody both reactive strategies and waypoint-oriented policies. This methodology has been validated on various robotic platforms. A cornerstone of their technique involves the use of hierarchical code generation, particularly by recursively elaborating on unspecified functions. This not only enables the generation of intricate code but also enhances the benchmark performance, achieving solutions for 39.8% of challenges on the HumanEval benchmark. In July 2023, Google released its Robotic Transformer 2 (RT-2) [17], which is a novel vision-language-action (VLA) model. RT-2 studies both online information and robot data, and turns this into basic instructions for controlling robots. The "language" part in VLA is driven by several LLM modeling including PaLI-X (55 billion of parameters), PaLI (3 billion of parameters), and PaLM-2 (12 billion of parameters). These models are co-fine-tuned with robotic trajectory data to output robot actions, which are represented as text tokens. They showed that their method produces highly effective robot policies and, crucially, offers improved adaptability and new abilities inherited from web-scale vision-language pretraining. They believe that this simple and general approach shows a promise of robotics directly benefiting from better vision-language models, which positions the domain of robot learning to capitalize on advancements from other fields. Building upon these advancements, the integration of LLM-driven robotic controls can profoundly transform robotics for built environment as well. By leveraging the vision-language-action models, these robots can be trained from vast amounts of environmental images (such as thermal imaging) and its correspondence actions under a low cost. It will greatly reduce the engineering cost and effort to develop robots for built environment. Given that building scientists are typically not experts in control or robotics, the capabilities of LLMs can bridge this knowledge gap, significantly accelerating the advancement and adoption of robots in the built environment.

**Interpretation of Machine Learning Control (MLC).** The potential of MLC [18] to enhance model-based control in HVAC systems is hindered by its opaque/black-box nature and inference mechanisms, which is challenging for users and modelers to fully comprehend, ultimately leading to a lack of trust in MLC decision-making. To address this challenge, interpretable machine learning (IML) has been widely applied in buildings [19]. However, majority of the existing IML literature pertaining to building applications leverage methodologies such as Shapley [20] for analytical exploration. While these studies quantify the importance and contributions of various inputs to their models, they lack a comprehensive and root-cause exploration of these findings in the context of specific control tasks, leaving a critical gap in the interpretation and understanding of the results.

LLMs hold significant potential in addressing this gap. They can be employed to undertake a deeper analysis of the results generated through IML studies in building applications. By utilizing tools like SHAP alongside LLMs, researchers can dissect the complex patterns and relationships in the data more effectively. The LLMs can aid in providing a richer understanding of the implications of various inputs on specific control tasks, thus facilitating a more comprehensive and nuanced analysis.

Furthermore, LLMs have the inherent capacity to synthesize information from a multitude of sources to create coherent and logically structured narratives. This characteristic stands them in good stead to address the challenges raised in the systematic arrangement of interpretations derived from multiple machine learning predictions in building control tasks. LLMs can be designed to aggregate and analyze the diverse sets of predictions, creating a "storyline" that logically connects individual predictive elements to offer a holistic view of the multi-step decision-making process. By doing so, they can help in constructing a rational pathway that delineates the series of control actions necessary based on the predictions. The development and integration of such systems would significantly enhance the transparency and effectiveness of MLC, steering them towards a more informed and optimized operational pathway.

## 3.2 Code Generation for Software and Modeling

LLM represents a significant leap in the realm of automated code generation. The inherent complexity of programming tasks often requires not only syntax correctness but also nuanced understanding and contextual reasoning – areas where LLMs excel due to their vast training on diverse datasets. LLMs can comprehend intricate problem statements in natural language and translate them into functional code snippets, reducing the manual effort and potential for human error. Furthermore, they can provide instant feedback, suggest optimizations, and even detect subtle bugs, streamlining the software development process. The adaptability of LLMs allows for seamless integration with various programming languages and frameworks, catering to a wide array of development needs. By bridging the gap between human intent and machine-executable code, LLMs are poised to significantly accelerate and enhance the code generation process, making software development more efficient and accessible. The application of LLM is widely discussed in general code development [21] and code repair [22]. Specifically in the field of building energy efficiency and decarbonization studies, LLM's capability in code generation can enhance the functionalities of tools (software, models, etc.), simplify user interaction, and streamline complex coding and modeling processes.

**Improve User Interaction**. LLMs have the potential to revolutionize user interactions by providing intuitive, natural language interfaces. By understanding and processing user inputs in a conversational manner, LLMs can offer more personalized and context-aware responses, eliminating the need for rigid command structures or extensive user manuals. This not only simplifies user experience but also empowers users, from novices to experts, to interact seamlessly with complex software, systems or databases. The end result is a more engaging and efficient interaction, fostering broader accessibility of tools and resources.

LLM can simply the users' interactions with professional software in this field, such as EnergyPlus, which requires years of users' experience in HVAC industry as well as energy modeling expertise. Zhang et al. [23] discussed potential applications of LLM in enhancing the interaction between user and building energy modeling tools. They specifically discussed how ChatGPT can enhance and streamline the BEM process, including 1) simulation input generation, 2) simulation error analysis, 3) visualization and interpretation of simulation output, 4) co-simulation, 5) simulation optimization, and 6) documentation improvement, including BEM training tutorial, knowledge interpretation, and extraction. This study illuminates the transformative potential of integrating LLM with BEM, improving its efficiency and accessibility across varied levels of modeling expertise.

**Speeding up Software Development.** Professional software is key to building energy efficiency and decarbonization. Building energy researchers and engineers often possess deep domain knowledge but may lack the programming expertise needed to create tailored software tools. Traditionally, they might rely on collaborations with software engineers, leading to prolonged development cycles due to the iterative back-and-forth communication required to refine the software. However, the advent of LLM is changing this landscape. With LLMs' proficiency in code generation and LLM-based tools like GitHub Copilot which already achieves great performance [24], these domain experts can now directly contribute to software development given either already written code or even natural language. Such tools, guided by the nuanced understanding of the researcher or engineer, can generate code, helping to bridge the technical divide. This not only accelerates software development but also ensures that the resulting tools are better aligned with the specific needs and nuances of building science.

**Streamlining Complex Data and Modeling Processes**. Modern energy studies are characterized by the integration of various systems they aim to analyze and optimize. Data is sourced from numerous areas—buildings, renewable energy sources, energy storage systems, power grids, human behaviors, transportation systems, and more. Each set of data represents a complex system that interacts with others, and human behaviors add another dimension of complexity. Traditional modeling approaches often struggle with the sheer volume and interconnected nature of this data. However, LLMs offer a promising solution. LLMs, with their capability for code generation, can process and organize these complex datasets efficiently, and further generate code tailored to specific modeling challenges. The most straightforward application is data-driven building energy prediction [25]. LLMs can streamline the process to analyze vast amounts of historical data, uncover complex patterns, visualize, train, and make accurate predictions about future trends of building energy consumption, providing a powerful tool for energy management and model predictive control.

### 3.3 Data Infrastructure

LLMs can play a significant role in improving the data infrastructure used in building energy efficiency and decarbonization studies. Given their ability to process, analyze, and interpret large volumes of data, LLMs can contribute to more effective data management, analysis, and utilization.

**Data Management and Organization**. Managing and organizing vast amounts of data is a key challenge in building energy efficiency and decarbonization studies. LLMs can assist in this process by automating data categorization and organization tasks. They can be trained to recognize different types of data and sort them into appropriate categories, making it easier for researchers and practitioners to find and use the data they need. For example, an LLM, when equipped with meticulously engineered prompts, can automatically categorize incoming data according to its type (such as energy consumption data, weather data, or occupancy data), its source (such as different sensors or systems), or its time period.

Navigating the abundant databases and open data sources in the building energy efficiency and decarbonization sector presents both an opportunity and a challenge. While this wealth of data is valuable, sifting through it to find relevant datasets is a daunting task. This is where LLMs come into play. LLMs can be integrated into user-friendly interfaces to help researchers quickly identify and access pertinent data with an advance technique called LLM-based semantic search [26]. By analyzing user input, these models can search and recommend datasets from various platforms that align with specific research needs based on LLM's trained understanding of users' description. This ensures that researchers can efficiently make use of the vast amount of available data, promoting more informed and data-driven decisions in the field.

**Data Integration and Schema Alignment**. Integrating different types of data and ensuring interoperability between different systems is another important aspect of data infrastructure. LLMs can assist in this process by interpreting data from different sources and converting it into another or common format. This can make it easier to combine and compare data from different systems, leading to more comprehensive and accurate analyses.

An LLM can be used to integrate energy consumption data from different sources such as load profile data from different utility companies. This involves not only aligning different data formats but also interpreting varying terminologies and measurement units used by each utility provider. This capability is particularly valuable in creating a unified view of energy usage across different geographical regions or consumer demographics. Zhang et al. [27] developed a semantic data schema for each open data category to maintain data consistency and improve model automation for Urban Building Energy Modeling. They also demonstrate that LLMs can automatically process open data based on the proposed schematic data structure. The accurate results generated by LLMs indicate that common open data can be easily aligned to another schema based on LLM. Zou et al. [28] highlighted the challenge of standardizing diverse building energy datasets. Current methods require time-consuming manual conversions of various building load profiles, with existing automated tools often falling short. They introduced DataTransformer, an innovative LLM-based approach, to address this gap. This approach combines a prompt template for capturing schemas, schema detection tools, prompt optimization, and a robust verification process. Their main contributions are: (1) an LLM-focused solution for building energy data standardization; (2) a benchmark dataset for real-world standardization challenges; (3) a thorough empirical assessment comparing DataTransformer to current tools. This method not only streamlines data transformation but also exemplifies the potential of LLMs in addressing specialized challenges, underscoring their role in promoting sustainable solutions.

An LLM can also convert data across standard semantic data models such as Project Haystack [29], Brick Schema [30], and ASHRAE 223P. The conversion is vital for integrating systems that are built on different semantic models, which is extremely important for ensuring interoperability and facilitating seamless communication between diverse building management systems. These models are crucial for ensuring that data related to building systems, energy management, and smart grid applications are interoperable and standardized. For instance, an LLM can take data structured in the Project Haystack format, which is widely used for tagging data in building automation systems, and convert it into the Brick schema, which is another semantic model focused on building and IoT data. The ability of LLMs to navigate these complex semantic models and facilitate data conversion is crucial for enabling seamless communication and data exchange between different systems in the smart building and energy sectors. However, as of the time of writing, no research papers have been identified that discuss this specific topic.

It needs further investigation across a wider range of diverse open data to further improve the data integration ability of LLMs. The key limitations are: 1) the ability of LLM to handle large documents and complicated schema mapping, e.g., many-to-one, aggregation, pivoting, transpose, and join, is not tested; 2) how to handle self-consistency issues (e.g., random responses) is not discussed. Finally, effective collaboration among researchers, industry experts, and policymakers is pivotal for the successful standardization of data for building energy studies. The trend is moving towards the development of open-source cyberinfrastructures where individuals can both publish and utilize open data based on a unified schema.

### 3.4 Technical Report and Paper Analysis

LLMs can analyze large volumes of textual data, identify patterns, extract valuable insights, and even generate human-like text. In the context of building energy efficiency and decarbonization studies, this ability can be leveraged in several ways.

**Analyzing Documents and Reports**. A substantial amount of information on building energy efficiency and decarbonization is found in textual reports. These could be technical reports, policy documents, or even news articles. LLMs can be trained to analyze these documents, extract key information, and present it in a structured and digestible format. For instance, an LLM can sift through a large number of energy audit reports, identify the most commonly reported energy inefficiencies, and present these trends to stakeholders. This can save countless hours of manual review and help stakeholders focus on the most critical issues. LLMs can also be employed to distill policy guidelines, focusing on primary goals, proposed strategies, and critical milestones. As an application, an LLM could extract essential directives from such policies, generating a visual timeline or scenario map that lays out the decarbonization journey over the next decade or more. This visualization would ensure alignment among a broad range of stakeholders, from construction magnates to local policymakers, creating a cohesive strategy to combat climate change through better building practices.

**Literature Interaction and Review Automation.** The field of building energy efficiency and decarbonization studies is rapidly evolving, and keeping up with the latest research from diverse disciplines can be a daunting task. LLMs can be trained to automatically understand new research papers, extract key findings, and summarize them. This can help researchers and practitioners interactive with the latest developments without having to read every single paper. While there are no existing papers specifically addressing the application of LLMs in automating literature interaction and review within the field of building energy studies, this topic has been explored in the context of extracting scientific text and papers in materials chemistry. This indicates a potential for cross-disciplinary adaptation and the opportunity to leverage the capabilities of LLMs in diverse scientific domains. Dunn et al. [31] demonstrated that "LLMs can effectively gather detailed scientific knowledge for three main tasks in materials chemistry: associating dopants with their respective host materials, indexing metal-organic frameworks, and extracting information related to general chemistry, phase, morphology, and application. This method offers an uncomplicated, readily available, and adaptable way to convert unstructured text into organized databases. This work also suggests the feasibility of applying LLMs in building energy studies. Besides, many LLM-based tools and plugins, such as "ScholarAI," are developed to help interactively connect users to peer-reviewed research articles.

This automation process not only condenses the essence of each paper into digestible summaries but can also categorize them based on relevance, impact, and novelty. Such categorization can guide researchers towards literature that aligns most closely with their specific interests or current projects. Furthermore, LLMs can cross-reference new findings with existing knowledge, highlighting possible synergies or contradictions, thereby enriching the depth and breadth of literature reviews. By automating the literature review process, not only is the research landscape made more accessible, but the rigor and comprehensiveness of reviews are also enhanced. The potential errors or biases of manual reviews can be minimized, ensuring a more objective and holistic understanding of the current state of the art in building energy efficiency and decarbonization studies.

### 3.5 Regulatory Compliance

The realm of building energy efficiency and decarbonization is governed by a myriad of regulatory standards, each dense with technical specifications, stipulations, and guidelines. For instance, standards like those set by ASHRAE, spanning numerous pages, detailing everything from building performance to HVAC specifications. Yet, their vastness is not the only challenge; these standards often undergo revisions, with updates emerging every few years. Keeping up with this dynamic landscape is a daunting task for stakeholders. Misinterpretations, lack of awareness of recent changes, or sheer oversight due to the overwhelming volume can result in non-compliance, with its consequent ramifications. In this complex and frequently updated regulatory environment, LLMs present a promising solution. While conventional NLP techniques have demonstrated some success in distilling complex regulatory documents [32], LLMs, with their advanced capabilities, have the potential to significantly elevate the quality and precision of interpretation. Given their capabilities to analyze and interpret large volumes of text, LLMs can be instrumental in helping stakeholders understand and adhere to these regulations and further automate the compliance process.

**Interpreting Regulations and Guidelines.** Navigating the intricacies of regulations and guidelines in building energy efficiency and decarbonization is not a straightforward endeavor. Their depth and technicality demand sophisticated tools for simplification and comprehension. For instance, an LLM could be employed to distill a comprehensive set of building energy regulations into a succinct summary, pinpointing pivotal requirements tailored for a specific project. This ensures stakeholders, even those without technical expertise, can quickly grasp the crux of the regulations, facilitating informed decision-making.

**Ensuring Compliance in Strategies and Plans.** The manual approach for checking building code compliance is labor-intensive, prone to human errors, and often incurs high costs. This limitation emphasizes the need for a more systematic and automated method, commonly termed Automated Compliance Checking (ACC). Numerous studies have ventured into ACC using conventional NLP methods. For example, Zhou et al. [33] combined NLP with context-free grammar to interpret intricate rules in ACC, enabling them to process regulatory text similarly to a domain-specific language. This study introduces a technique and structure for automated rule interpretation with broad applicability, aiming to convert diverse textual regulatory documents into computable rules. Additionally, the study provides an inaugural dataset for regulations, facilitating further investigations, validations, and benchmarks in the ARC domain. Wu et al. [34] introduced a framework aiming for maximum automation with minimal user input, called the Invariant signature, logic reasoning, and Semantic NLP-based Automated Building Code Compliance Checking framework. Using this structure, they crafted an ACC prototype and evaluated it using Chapter 10 of the International Building Codes 2015, attaining a precision of 95.2% and a recall of 100% in detecting non-compliance across two projects. Zhang and El-Gohary [35] introduced a rule-driven NLP technique to autonomously pull requirements from regulatory texts for compliance verification. This method blends pattern recognition with conflict resolution rules, tapping into both syntactic and semantic text aspects. They made use of phrasal tags based on Phrase structure grammar and a strategy for sequencing and segregating semantic details to minimize pattern requirements. When assessed using the 2009 International Building Code, their method attained a 96.9% precision rate and a 94.4% recall rate.

While conventional NLP methods exhibit proficiency in numerous applications, they face inherent challenges when interpreting ambiguous or intricate regulatory text. Notwithstanding these systems achieve impressive metrics, they often require extensive domain-specific customization and can be challenged by evolving regulatory versions or variations in language and terminology. The advent of LLMs presents opportunities for improving ACC further. Due to their advanced architectures and ability to process large amounts of data, LLMs might provide a more in-depth and detailed compliance assessment compared to traditional NLP models. Their capabilities allow for a broader interpretation of regulations, which could enhance the accuracy and reliability of the compliance process. By analyzing the details of a plan and comparing them with regulatory requirements, an LLM not only can identify potential compliance discrepancies, but also can recommend appropriate modifications.

As LLMs represent a cutting-edge progression in the field, scholarly endeavors are underway to evaluate their relevance and performance in ACC. Recently, Liu et al. [36] introduced an early application of GPT models for ACC without the need for supplementary domain knowledge or terminology clarification. Their approach consists of directly inputting building design specifications and associated codes into the model, steered by a specific task prompt. Subsequently, the GPT models produce compliance outcomes. Preliminary tests on an artificially generated dataset yielded up to 91% accuracy, demonstrating the model's effectiveness.

## 3.6 Building Lifecycle Management

The construction and maintenance of buildings is an intricate and collaborative process involving multiple stakeholders, each with their own unique roles and responsibilities. From the designer's vision to the contractor's practical application to the maintainer who ultimately ensures the building's longevity, the lifecycle of a building encompasses countless interactions, decisions, and communications.

Effective communication among the myriad stakeholders is critical and challenging, especially given a large amount of documents that are generated. Various studies have investigated conventional NLP to automate documentation in building management. For example, in construction reporting, the extraction of reporting requirements from extensive construction contracts is frequently done by hand, owing to a deficiency in appropriate methodologies. The manual approach frequently leads to rough, often underestimated, calculations of the time and expenses associated with fulfilling these requirements. Without a precise comprehension of the stipulations, contractors are hindered from making enhancements to reporting processes before the commencement of the project. This informal and potentially inaccurate approximation can stifle efficiency and innovation within the workflow. To overcome such challenge in construction reporting, Jafari et al. [37] formulated a framework for automatically identifying reporting requirements and forecasting time and costs. Leveraging NLP and ML techniques, they discerned reporting stipulations from contract documents and employed probabilistic simulations to estimate the overhead expenses and timeframes tied to report creation.

In the context of building maintenance, document-driven tasks like staff assignment are traditionally conducted manually, consuming significant time and effort. Innovative solutions have emerged to counteract this inefficiency. For example, Pan et al. [38] used NLP and ML approaches to read service request and assign workforce and priority automatically. Their model, developed and tested using 3-year worth maintenance records from 60 building on a university campus, yielded 77% and 88% accuracy in workforce and priority predictions, respectively. Similarly, Bouabdallaoui et al. [39] developed a ML algorithm based on NLP to classify maintenance requests. Developed and tested using 10-year worth maintenance records from a healthcare facility, their model yielded an average accuracy of 78% in classifying work requests.

The successes achieved using conventional NLP offer a glimpse into the transformative potential of using LLMs. With their expansive training data and sophisticated architectures, LLMs promise not only to streamline document extraction and interpretation but also to understand nuances and context better. Their advanced capabilities may revolutionize stakeholder communication, streamline operations, and ultimately lead to smarter and more sustainable building practices.

## 3.7 Education and Training

The transformation of the built environment towards greater energy efficiency and decarbonization is contingent upon the proficiency of professionals steering these advancements. This necessitates a rigorous, specialized education and training paradigm. Recognizing this imperative, Recently, ASHRAE, alongside its collaborative partners, secured a grant of $2.85 million from the U.S. Department of Energy Building Technologies Office, designated for the Resilient and Efficient Codes Implementation (RECI) initiative [40]. This initiative, named "Energy Code Official - Training & Education Collaborative" (ECO-TEC), seeks to enhance energy code enforcement. The anticipated outcomes include substantial financial savings, reaching up to $335 million by the fifth year, and significant reductions in $CO_2$ emissions. This commitment at the federal level underscores the pressing need for targeted educational and training programs in building energy efficiency and decarbonization.

In this context, LLMs emerge as a potent tool in enhancing educational materials and training programs related to building energy efficiency and decarbonization. With the capability to generate human-like text and interpret intricate details, LLMs present opportunities to design engaging and relevant learning resources that resonate with modern professionals and educators. In the following subsections, we will delve into the potential roles that LLMs can play in enhancing and streamlining educational and training endeavors.

**Developing Educational Materials**. LLMs can be used to develop a wide range of educational materials on building energy efficiency and decarbonization. For instance, they can generate easy-to-understand explanations of complex concepts, create interactive quizzes, or even develop comprehensive textbooks. For example, an LLM could be used to generate a series of articles explaining different energy efficiency measures in a way that is accessible to a general audience.

**Creating Interactive Training Programs**. LLMs can also be used to create interactive training programs for professionals in the field of building energy efficiency and decarbonization. These programs could include interactive modules where users can ask questions and receive answers in real time, or scenario-based training where users can learn how to make decisions in simulated real-world situations. For instance, an LLM could be used to develop a training module where building engineers can learn about the latest decarbonization strategies and test their knowledge through interactive scenarios.

**Continuous Learning and Updating Knowledge**. Given the rapid advancements in building energy efficiency and decarbonization studies, continuous learning is crucial for professionals in this field. LLMs can support this by regularly updating training materials based on the latest research and developments. For example, an LLM could be used to automate the process of updating a course on building energy management with the latest research findings and best practices.

While LLMs show tremendous potential in reshaping education related to building energy efficiency and decarbonization, their impact extends to other educational domains as well. To gain a comprehensive view of their transformative power across educational landscapes, it is instructive to reference the work of Kasneci et al. [5]. This study delves into the benefits and hurdles of using LLMs in education from the perspectives of educators and students. It highlights the potential of these models in creating educational materials, enhancing student participation, and delivering tailored learning experiences. The research underscores the importance of educators and students understanding the strengths and constraints of these tools. Integrating these models into curricula requires a focus on critical thinking and verifying information. Although challenges like biased outputs, the essential role of human oversight, and potential misuse arise in AI contexts, they can be managed. The piece wraps up with recommendations for addressing these issues ethically and responsibly within the educational realm.

# 4 Challenges and Limitations

In the preceding section, we explored the capabilities of LLMs. Here, rather than emphasizing their strengths, we will delve into their limitations. Specifically, we will address the challenges and constraints of deploying LLMs in the fields listed in Section 3.

## 4.1 Complex and Expensive Computation

LLMs are complex models that require significant computational resources for training, fine-tuning, and inference. This can lead to high energy consumption and carbon emissions, which is counterproductive in the context of building energy efficiency and decarbonization studies.

The financial implications of deploying LLMs can be substantial. Relying on API-based solutions often comes with associated costs, especially when processing large volumes of data or requiring frequent access. On the other hand, running LLMs on local infrastructure necessitates investment in high-end computational hardware, particularly Graphics Processing Units, which are essential for efficient model operations. The upfront costs of these hardware components can be considerable, and they may require periodic upgrades to keep pace with advancing model complexities.

Technological advancements in software and hardware are driving the development of more efficient and cost-effective LLMs such as Llama-2 7B version. As these models become more prevalent, economies of scale and continuous innovations are expected to make them increasingly accessible for diverse applications.

## 4.2 Data Privacy, Security, and Copyright

Building energy efficiency and decarbonization studies often rely on multiple information and data sources. One of the primary challenges in using LLMs in building energy efficiency and decarbonization studies is ensuring data privacy and security. LLMs are typically trained on large amounts of data, which may include sensitive information. There is a risk that such sensitive information could be unintentionally revealed during the model's output, even if it is not directly related to the input. Moreover, the data used for training and operations could be vulnerable to unauthorized access or cyber-attacks, leading to potential privacy breaches. Therefore, strict data privacy and security measures must be in place when using LLMs.

When using LLMs to generate content and solutions, there also exists the possibility of inadvertently reproducing copyrighted material, leading to intricate intellectual property issues. A substantial portion of academic literature, technical reports, and building code standards, including those from respected entities like ASHRAE, are copyrighted. Accessing many of these resources requires purchase or specific permissions. Training or fine-tuning LLMs using such copyrighted materials might result in the generation of content that is not intended for public viewing without purchase or alteration without authorization. As such, both developers and users of LLMs must exercise caution when leveraging such materials, ensuring adherence to intellectual property rights and licensing agreements.

The matter of ownership regarding insights or content exclusively produced by an LLM remains ambiguous, necessitating further legal exploration and clarity. While rectifying this issue will undoubtedly demand significant effort from lawmakers, authors in the building industry must also remain vigilant. They should actively consider the implications of how their original work might be used or replicated.

## 4.3 Complexity of Fine-Tuned LLMs: Training Data Availability and Model Uncertainty

Within the realm of building energy efficiency and decarbonization, it is evident that several prominent LLMs, such as ChatGPT and Llama, have not been extensively trained on specialized knowledge from this domain. This shortcoming can impede the direct applicability of these models to real-world challenges in the sector. The key avenue to address the limitation of foundation model is through fine-tuning, where the model is further trained on domain-specific datasets to enhance its accuracy and relevance.

The efficacy of fine-tuned LLMs hinges largely on the quantity and quality of their training data. A deficit in pertinent, high-caliber data or corpora can hinder LLMs from producing precise predictions or extracting significant insights. There is a notable absence of a comprehensive and well-curated corpora tailored for fine-tuning and semantic search in the realm of building energy efficiency and decarbonization. This gap stems from the multifaceted and interdisciplinary nature of the field. There is a pressing need for an aggregated, reusable corpus that encompasses text and data on building carbon emissions, including knowledge from building science, construction material information, occupant behavior patterns, building energy metrics, building regulations, energy policies, and technical reports, etc.

While fine-tuning can indeed enhance the model's understanding and proficiency in this particular domain, it can simultaneously lead to an unforeseen decline in performance in other (often unrelated) areas especially the in-context learning ability [41]. This decline is not always predictable or easy to track, resulting in a tool that, while more specialized, is potentially less reliable for a broad array of tasks. Moreover, this specialization can sometimes lead to outputs that are not just less accurate but diverge significantly from expected results, creating a new set of challenges in model utilization and interpretation. Therefore, as we move forward, it is vital to develop strategies for fine-tuning that retain a model's general applicability while enhancing its domain-specific performance. This necessitates ongoing research into the underlying mechanisms governing LLM behavior to strike a balance between specialization and generalization, ensuring reliable and consistent performance.

Fine-tuning is not the only way to develop specialized LLMs. An alternative approach is to employ LLM-based semantic search, a method that leverages the model to mine relevant domain-specific information from external knowledge bases [42]. This strategy involves integrating domain expertise not presented in the initial training by utilizing semantic search methods, thereby aligning the LLMs more closely with the intricacies of the building energy field and enhancing their performance in specialized tasks. The advantage of LLM-based semantic search over fine-tuning lies in its flexibility, cost-effectiveness, and the greater degree of certainty it can offer. While fine-tuning requires significant investments in computational resources and access to high-quality, labeled datasets, semantic search is more economical, tapping into external sources of knowledge to focus the model's expertise more precisely. This strategy is not only cost-saving but also flexible, adapting easily to the nuanced requirements of specific domains. Furthermore, it fosters a higher level of certainty in the model's performance, steering clear of the unpredictable behavioral fluctuations frequently seen in fine-tuned models, and ensuring a stable and dependable tool for specialized tasks in the building energy efficiency and decarbonization sector.

### 4.4 Self-Consistency

LLMs grapple with uncertainties in their computational outcomes, often related to self-consistency issues, such as producing random responses. Professionals in the building energy sector must remain vigilant and account for these potential inconsistencies when leveraging LLMs.

Researchers are continuously exploring avenues to foundationally enhance the self-consistency of LLMs. Wang et al. [43] introduced a self-consistency approach, suggesting that difficult problems often have several perspectives but converge to one right answer. While such theoretical advancements lay the foundation for more consistent LLMs, practical strategies are equally essential: strategies such as continuous validation, cross-referencing with other data sources, and expert reviews can help address the challenges posed by LLM-induced uncertainties.

# 5 Future Research Directions

## 5.1 Enhancing LLMs for Domain-Specific Tasks

Future research can focus on enhancing LLMs for domain-specific tasks in building energy efficiency and decarbonization studies. While LLMs are already capable of processing and analyzing large amounts of data, further improvements can be made to tailor them to specific tasks in this field.

An essential trajectory for forthcoming research centers around delineating the most suitable approach among fine-tuning, prompt engineering, or semantic search for distinct tasks. Rather than universally applying a single method, the optimal strategy may vary based on the specific requirements and nuances of each challenge. For instance, tasks demanding deep domain knowledge might benefit most from fine-tuning, where the model is further refined on specialized datasets. Wu et al. [11] developed BloombergGPT, the first specialized LLM for the financial domain. BloombergGPT is a language model with 50 billion parameters, trained using an extensive variety of financial data. The concept of creating a similar model such as "BuildingEnergyGPT" is intriguing. However, the contexts in which we deploy it warrant thoughtful consideration. In contrast, scenarios requiring a broader understanding without deep specialization might be aptly tackled through prompt engineering, optimizing the way LLMs interact with and interpret a given question or data set. Meanwhile, when the priority is swiftly locating relevant insights within vast data reservoirs, semantic search could emerge as the most efficacious tool. In essence, discerning the best-fit method for each task will be crucial, not only to maximize the efficiency and accuracy of LLMs but also to ensure that they remain versatile and adaptive across a spectrum of challenges.

Future directions in Enhancing LLMs for domain-specific tasks in this field include: 1) **Specialized Training Datasets for LLM Fine-Tuning:** Curate and develop domain-specific datasets enriched with building energy efficiency and decarbonization knowledge. These datasets can be used to fine-tune models like "BuildingEnergyGPT" to enhance their specificity and relevance, 2) **Prompt Engineering Research**: Delve deeper into prompt engineering to optimize LLM interactions for domain-specific tasks, thereby ensuring that questions or tasks are interpreted with higher accuracy, self-consistency, and context-awareness, and 3) **Semantic Search Enhancements**: As we acknowledge the potential of semantic search in quickly locating relevant insights, future research can focus on improving its efficiency and precision, especially in domain-specific contexts.

## 5.2 Multi-Modal LLMs

Multi-modal LLMs represent a frontier in AI technology, where the systems are designed to process and integrate information from various types of data — including, but not limited to, text, images, and videos — to perform more complex and nuanced tasks. These models leverage the strengths of individual AI technologies, such as NLP and computer vision, to create a synergistic and more capable system.

In the building sector, the deployment of multi-modal LLMs opens up a promising landscape for innovation, offering comprehensive and effective solutions by amalgamating LLMs with other AI technologies. This multi-dimensional approach can significantly enhance the analysis and interpretation of a diverse array of data types prevalent in the sector, fostering a richer understanding and facilitating intelligent automation in various processes. For instance, integrating LLMs with computer vision technologies can enable the simultaneous analysis of visual data such as building blueprints or thermal imagery alongside textual data, providing a deeper, more holistic view of building systems and environments. Moreover, it can aid in the development of robotic control systems for built environments based on real-time feedback from videos and images, thereby enhancing operational efficiency and responsiveness.

Such a synergistic approach could lead to a plethora of applications including, but not limited to, HVAC design assistance and automated HVAC diagram generation, leveraging the combined strengths of visual and textual analysis to handle complex, domain-specific tasks more effectively and efficiently.

### 5.3 Collaborative Research Between AI and Energy Experts

Lastly, fostering collaborative research between AI and energy experts can lead to more effective use of LLMs in building energy efficiency and decarbonization studies. Such collaborations can help bridge the gap between the capabilities of LLMs and the specific needs and challenges of this field. Energy experts can provide domain-specific knowledge and insights to guide the development and application of LLMs, while AI experts can bring their technical expertise to bear on complex data analysis and modeling tasks. Energy experts, representing the "demand side," guide collaborations based on their understanding of the pressing issues at hand. Meanwhile, AI specialists offer the technical prowess to devise solutions.

However, it is noteworthy that a large portion of AI expertise is currently drawn towards booming industries such as medical science, commerce. Hence, establishing a mechanism to steer the attention and contributions of AI experts towards the building sector is indispensable. Identifying incentives and creating awareness about the pressing issues and the potential impact of AI in this sector can be pivotal. For this interdisciplinary endeavor to thrive and evolve, a close-knit collaboration between the two is paramount, underpinned by a conscious effort to attract AI talents to focus on building energy efficiency and decarbonization.